\documentclass{elsart3p}
\usepackage{graphicx,amssymb}
\usepackage{subfigure}
\usepackage{dcolumn}
\usepackage{bm}

\begin{document}
\begin{frontmatter}
\title{Work Fluctuations and Stochastic Resonance}%
\author{Shantu Saikia}{$^1$}, %
\author{Ratnadeep Roy}{$^2$}, %
\author{A.M.~Jayannavar{$^3$}\corauthref{cor}}%
\corauth[cor]{Corresponding Author}%
\ead{jayan@iopb.res.in}%
\address{{$^1$}St.Anthony's College, Shillong-793003, India}
\address{{$^2$}Women's College, Shillong-793003, India}
\address{{$^3$}Institute of Physics, Sachivalaya Marg, Bhubaneswar-751005, 
India}

\begin{abstract}
We study  
Brownian particle motion in a double-well potential driven by an ac force. 
This 
system exhibits the  phenomenon of stochastic resonance. Distribution 
of work done on the system over a drive period in the time asymptotic 
regime have been 
calculated. We show that fluctuations in the input energy or work 
done dominate the mean value. The mean value of work done over a period as
a function of noise strength can also be used to characterise stochastic 
resonance in the system. We also 
discuss the validity of steady state fluctuation theorems in this 
particular system.
\end{abstract}


\end{frontmatter}

\section{Introduction :}
Stochastic resonance (SR) refers to the enhanced response 
of a system to a small deterministic periodic forcing in the presence of an 
optimal amount of noise \cite{Gammaitoni}. It often occurs in bistable 
and excitable systems 
with subthreshold inputs. Noise plays a constructive role in this
phenomenon.This is due to 
a cooperative interplay between nonlinearity of the system, input signal 
and noise.
Thus the power from 
the whole spectrum is pumped into a single mode that is coherent with
the external driving force. Because of its generic nature, this phenomenon,
boasts applications in almost 
all areas of natural sciences. Different quantifiers 
of SR have been discussed\cite{Dan2,Iwai,Dan3,Riemann,Giterman}
as regards to its validity as 
a bonafide resonance, i.e., the resonance phenomena as a 
function of noise strength as well as the driving frequency.  
Hysteresis loop area\cite{Dan2}, input energy or the work done on 
the system per cycle\cite{Iwai,Dan3} and area under the first peak in 
the residence 
time distribution\cite{Dan1} have turned out to be good quantifiers 
characterizing SR as a bonafide resonance.
Recently it has been shown that the different quantifiers of SR are
mathematically  
related to each other\cite{Riemann}. Earlier studies on the work done in a 
periodically driven
bistable system by an external agent 
has established that 
the average work peaks around the resonance, as expected
\cite{Iwai,Dan3}. In
our present work, we study the fluctuations and nature of the 
probability distributions  of work done over a period in a driven 
double-well system in the time 
asymptotic
regime (where all averaged quantities become periodic in time with a period 
of the external drive). 

Motivation for this study 
also comes from recent interest in the so called fluctuation theorems and 
transient violation of second law in systems driven to a nonequilibrium 
state\cite{Bustamante,Ritort}.
Fluctuation theorems describe properties of the distribution of various 
nonequilibrium quantities, such as work, heat, entropy, etc.
\cite{Bustamante,Ritort,Evans}. It is 
important to note that these theorems are rigorous and applicable to 
systems driven arbitrarily far away from equilibrium. These theorems can be 
extremely useful in analysing the role of fluctuations on the performance
characteristics of engines at nanoscales (e.g., molecular motors).
In these tiny systems interactions with the environment
are dominated by thermal fluctuations\cite{Bustamante}. Moreover 
fluctuations in physical 
quantities are more than the mean values and large fluctuations may occur
which can lead to unexpected consequences. It has been shown recently 
that overdamped driven harmonic oscillator (linear system) 
satisfies a steady
state fluctuation theorem (SSFT) for the work done over a single period 
($W_{p}$) in the time asymptotic regime\cite{Mamata1}. 
SSFT implies that the 
probability distribution $P(W_p)$ of work $W_{p}$ satisfies the following 
relation\cite{Evans,Mamata1,Cohen,Jarzynski,Douarche}, namely,

\begin{eqnarray}
\frac{P(W_{p})}{P(-W_{p})}=e^{({\beta W_{p})}}.
\end{eqnarray}

We show that this is an artifact of the linearity of the model and in 
general does not hold for driven nonlinear systems. However, we show 
that SSFT for the work is indeed satisfied, if one instead 
considers the work done over a large number of cycles\cite{Douarche}. 
Our study of input energy 
fluctuations also reveals that the relative variance is larger than one across 
SR indicating that fluctuations dominate the mean value.

\section{The Model:}
\begin{figure}[t]
\includegraphics[width=7cm,height=8cm,]{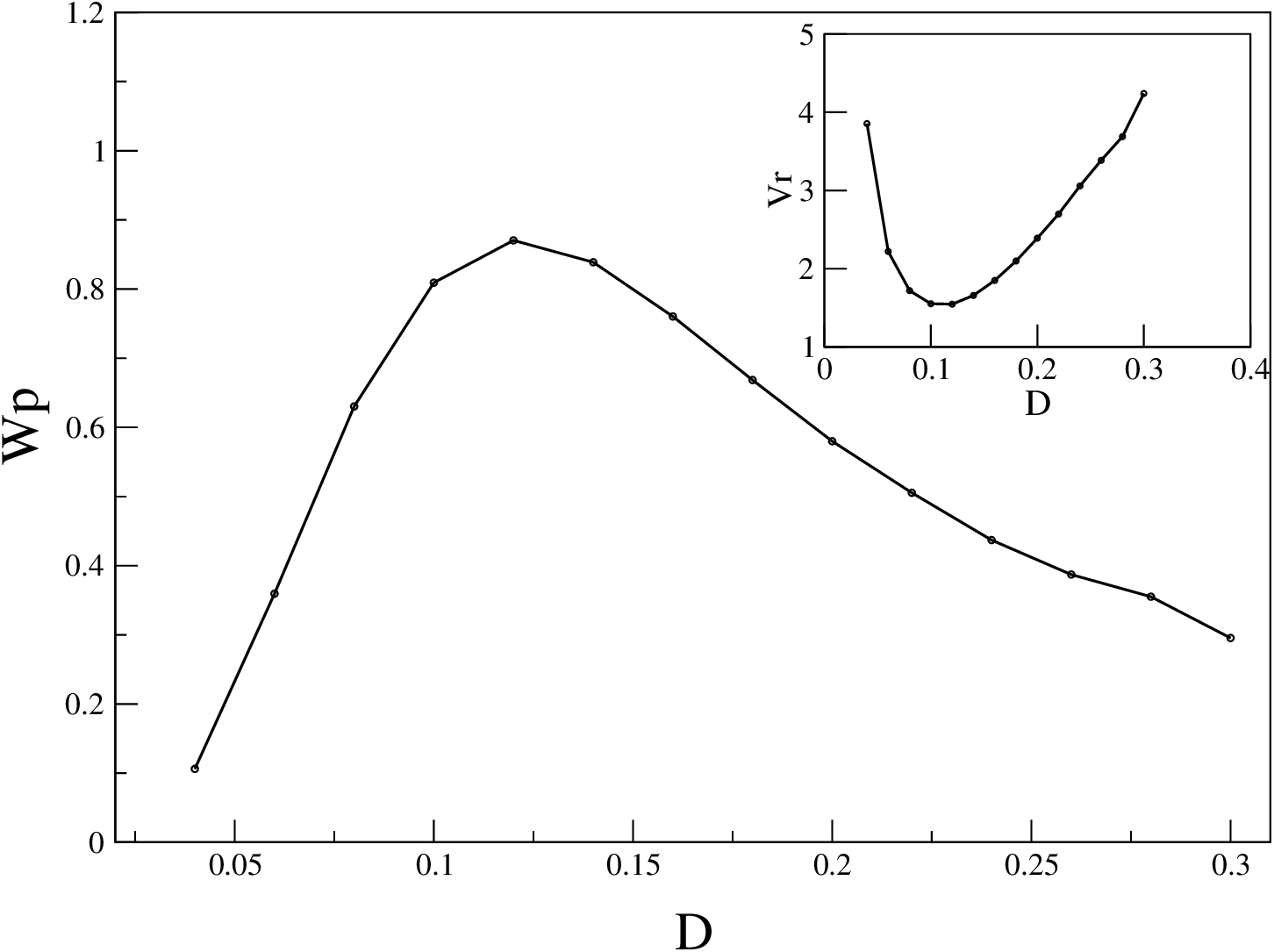}
\caption{\label{fig:epsart}Figure shows the variation of average work $W_p$
done per period with temperature D. Inset shows the corresponding plot
of relative variance $V_r$. The parameters are A=0.1, and $\omega=0.1$.}
\end{figure}

\begin{figure*}[ht]
  \centering
  \subfigure[]{\label{fig:edge-b}\includegraphics[width=7.5cm,height=6cm]{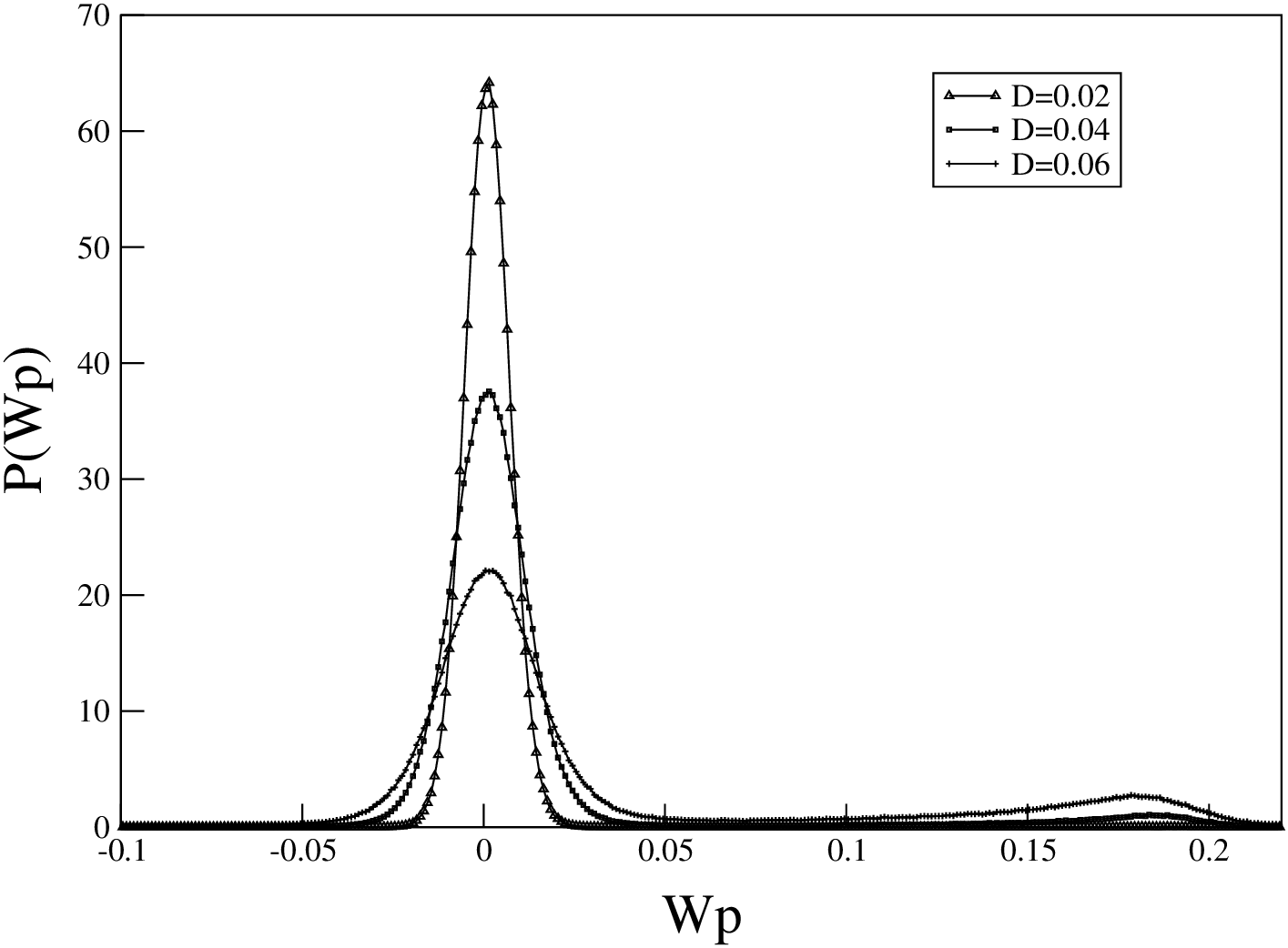}}
\hspace{0.4cm}
\subfigure[]{\label{fig:edge-c}\includegraphics[width=7.5cm,height=6cm]{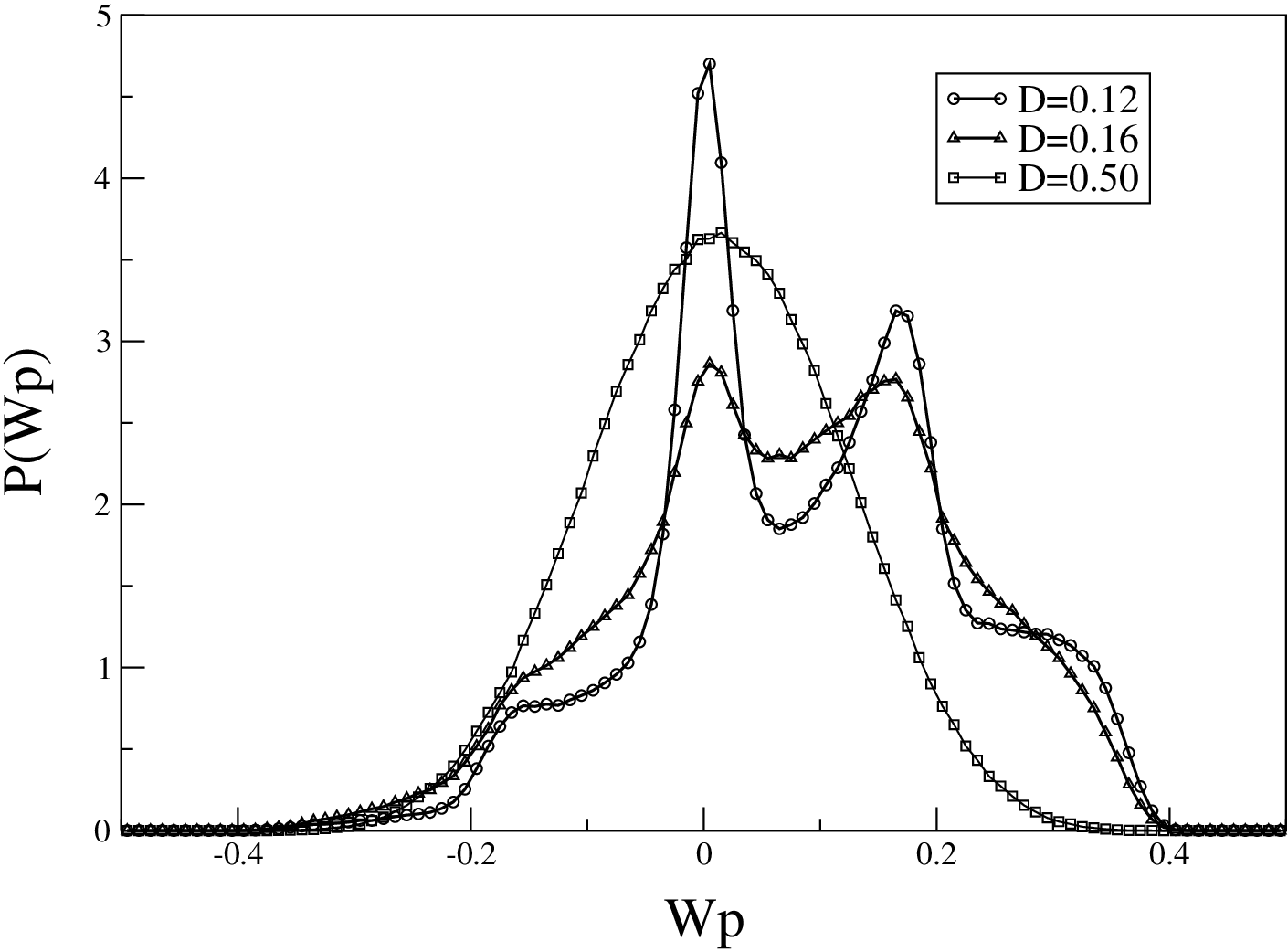}}
  \caption{Probability distribution for various values of D, other parameters being the same as in Fig.1.}
\label{fig:edge}
\end{figure*}
\vspace{1cm}

\begin{figure}[htp]
  \subfigure[]{\label{fig:edge-d}\includegraphics[width=6cm,height=5cm]{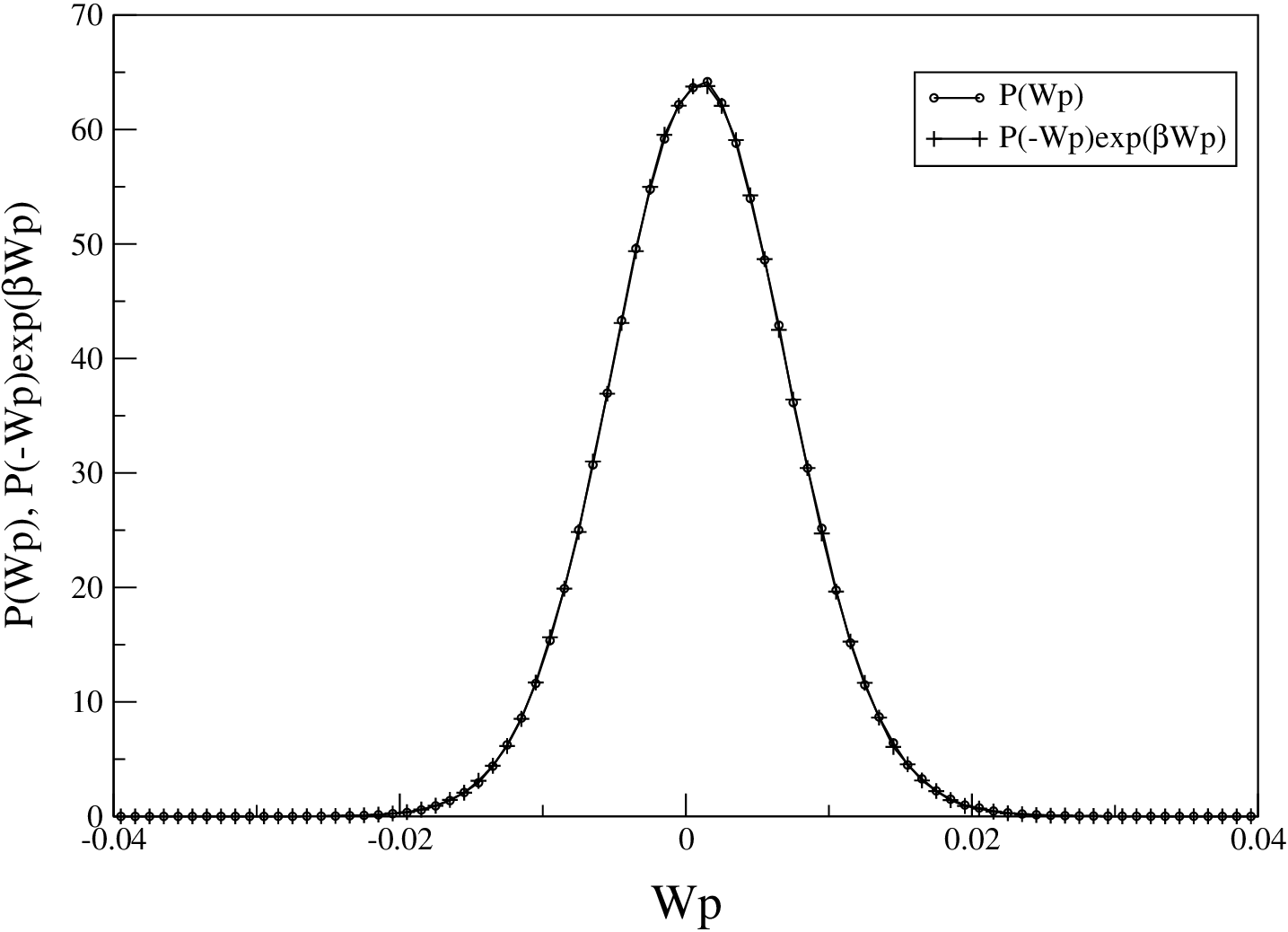}}\\
\subfigure[]{\label{fig:edge-e}\includegraphics[width=6cm,height=5cm]{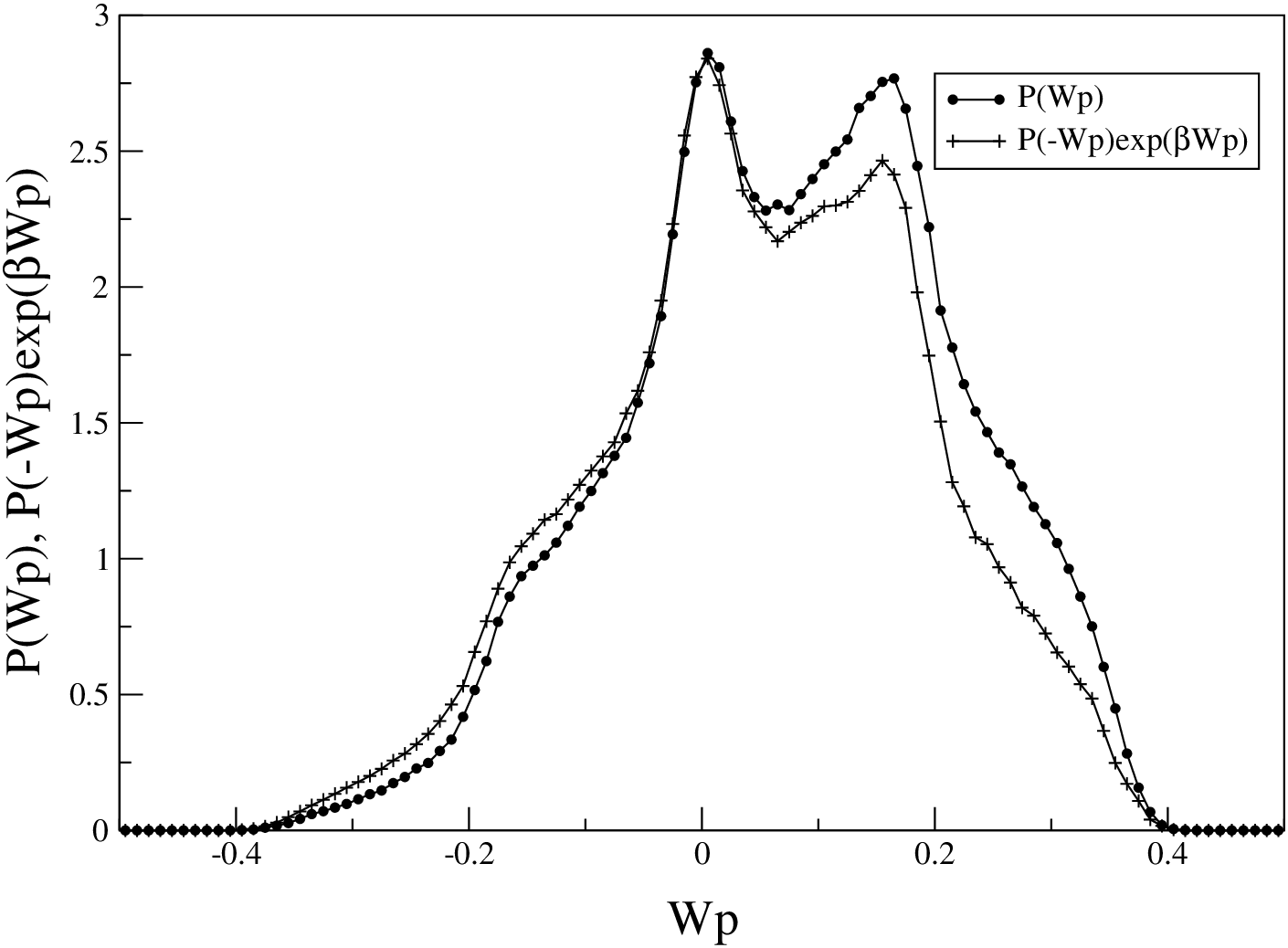}}\\
\vspace{0.6cm}
\subfigure[]{\label{fig:edge-f}\includegraphics[width=6cm,height=5cm]{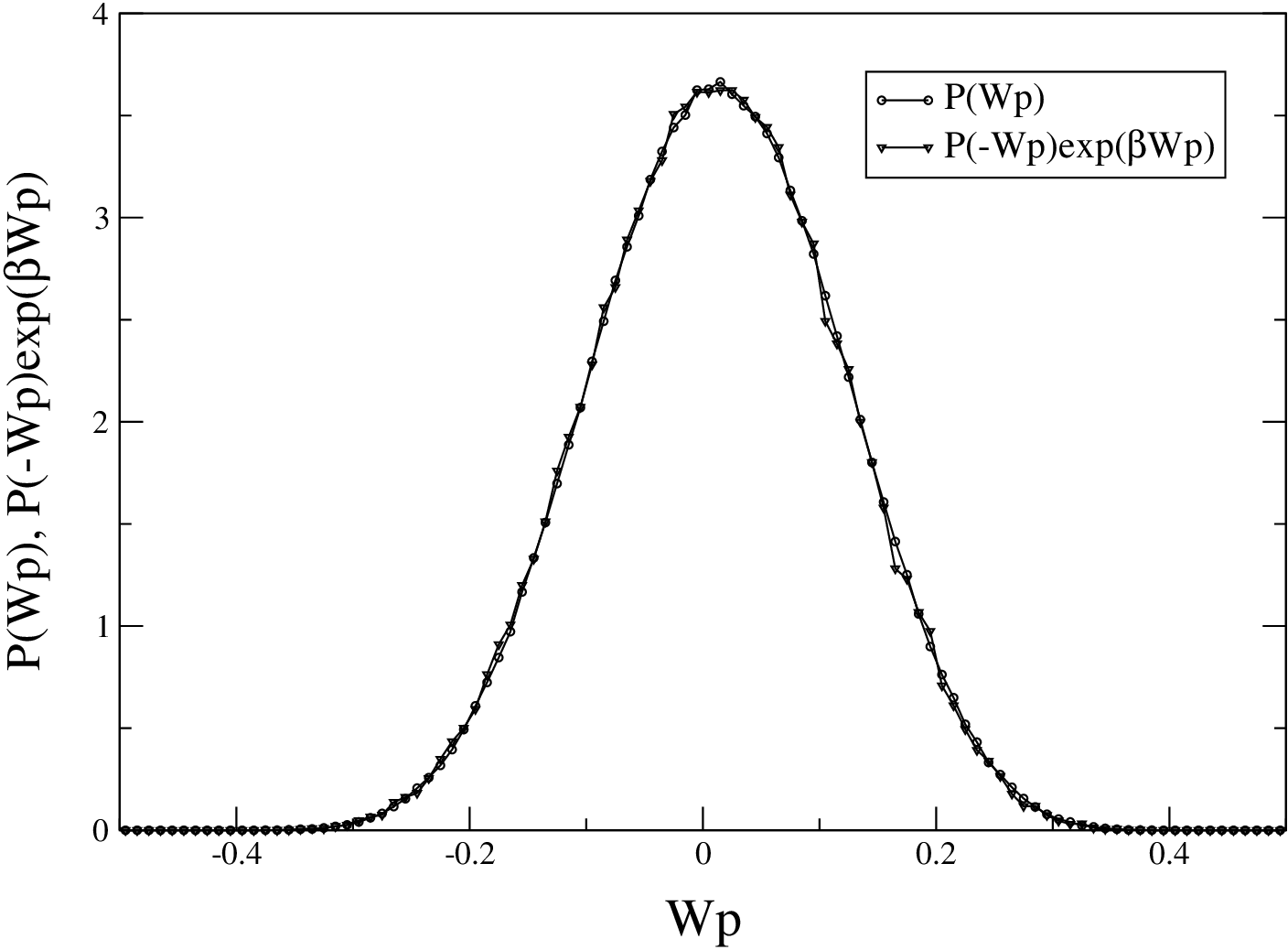}}
  \caption{}{Plot of $P(W_p)$ (line with circles),
$P(-W_p)exp(\beta W_p)$(line with crosses) vs. $W_p$.
Plots are for D=0.02(a), D=0.16(b)  and D=0.5(c)
respectively. For all plots $\omega=0.1$ and $A=0.1$. }
\label{fig:edge}
\end{figure}

The overdamped Langevin equation of a particle in a double-well potential 
in the presence of a time periodic force is given by\cite{Cohen} 
\begin{eqnarray}
 \frac{dx}{dt}=-\frac{\partial{U(x)}}{\partial{x}} + \xi{(t)}, 
\end{eqnarray}
where 
\begin{eqnarray}
U(x)=\frac{x^{4}}{4}-\frac{x^{2}}{2}-A x\sin\omega t.
\end{eqnarray}
We have set friction coefficient ($\gamma$) to unity. The random force field 
$\xi{(t)}$ is a zero mean Gaussian white noise, i.e,
\begin{eqnarray}
\langle\xi{(t)}\xi{(t^{'})}\rangle=D_{0} \delta{(t-t^{'})}. 
\end{eqnarray}
where $D_{0}=2\gamma k_{B}T$ is the strength of the noise. The static 
double-well potential $\displaystyle V(x)=\frac{x^{4}}{4}-\frac{x^{2}}{2}$, 
has a barrier height
$\Delta{V}=0.25$ between the two symmetrical wells (minima) located at 
distances $x_{m}=\pm 1$. We consider the case of weak forcing 
$A|x_{m}|<\Delta V$.All the physical quantities are taken in dimensionless 
units as prescribed in references\cite{Iwai,Dan3}.
\vspace{.6cm}

The work done by the external drive on the system or the input energy
$\displaystyle\left(\tau_{\omega}=\frac{2\pi}{\omega}\right)$ is defined
as\cite{Iwai,Dan3}
\begin{eqnarray}
W_{p} &=& \int^{t_0+\tau_{\omega}}_{t_0}\frac{\partial{U(x,t)}}
{\partial{t}}\,dt,\nonumber\\
    &=& -A\omega\int^{t_{0}+\tau_{\omega}}_{t_0} x(t)\cos\omega t\,dt
\end{eqnarray}
This follows from the stochastic energetics formalism developed by 
Sekimoto\cite{Sekimoto}.

Numerical simulation of this model was carried out by using Huen's method
\cite{Huen}. To
calculate the work done we first evolve the system and neglect initial 
transients and then work done over a period ($W_p$) is calculated (eqn.(5)). 
The values of $W_p$ are different for different cycles. To get better 
statistics 
we have calculated $W_p$ for 100000 different (sometimes even more) cycles. 

\section{Results and Discussion:}
In Fig.1 we plot the average work done $(W_p)$ over a single period
(in the time asymptotic regime) as a function of noise strength D($=kT=
\frac{1}{\beta}$, in
dimensionless units) for
low amplitude driving of strength $A=0.1$ and for frequency $\omega=0.1$.
We clearly see that input energy exhibits a peak as a function of D.
This can be attributed to the synchronized escape from the potential well
with the external periodic drive. We reproduce exactly the same figure
as in reference\cite{Dan3}, which has been obtained using a different
numerical method.

For the case where noise is small the particle remains in one of the potential
wells for a longer time(intrawell motion dominates) with an occassional random 
jump to another well (Kramers' escape over the barrier) as a function of time.
When the noise is very strong a large  number of interwell 
random switches occur for 
each period
of the sinusoid and the systems response is again random 
(interwell motion dominates).
In between these two conditions, surprisingly, there exists an optimal 
value of the noise that cooperatively concurs with the periodic forcing 
to make almost exactly one switch per period. Hence interwell motion will 
be in synchrony with the input signal. Quantitatively this condition is 
determined by the matching of the two time scales \cite{Gammaitoni,Dan3}, 
namely, the period of the 
input signal and the Kramers' escape rate $r_k$, 
$\displaystyle\left(\frac{1}{r_k}=\frac{\tau_\omega}{2}
\rightarrow\frac{\omega}{\pi}=r_k, {\textrm{with}} ~ 
r_k=\frac{1}{2\pi}e^{(\frac{0.25)}{D})}\right)$. 
At 
the optimal value of noise the input energy is maximum. The nature of this 
maximum is a function of other physical parameters as discussed in detail
in\cite{Iwai,Dan3}.
In the inset of Fig.1 we have plotted the relative variance 
$\displaystyle V_r=\sqrt\frac{\langle {W_p}^2\rangle - \langle {W_p}\rangle^2}
{\langle{W_p}\rangle^2}$    
of work done $W_p$ over a period as a function of 
noise strength. 

\begin{figure}[htb]
\includegraphics[width=7cm,height=6cm]{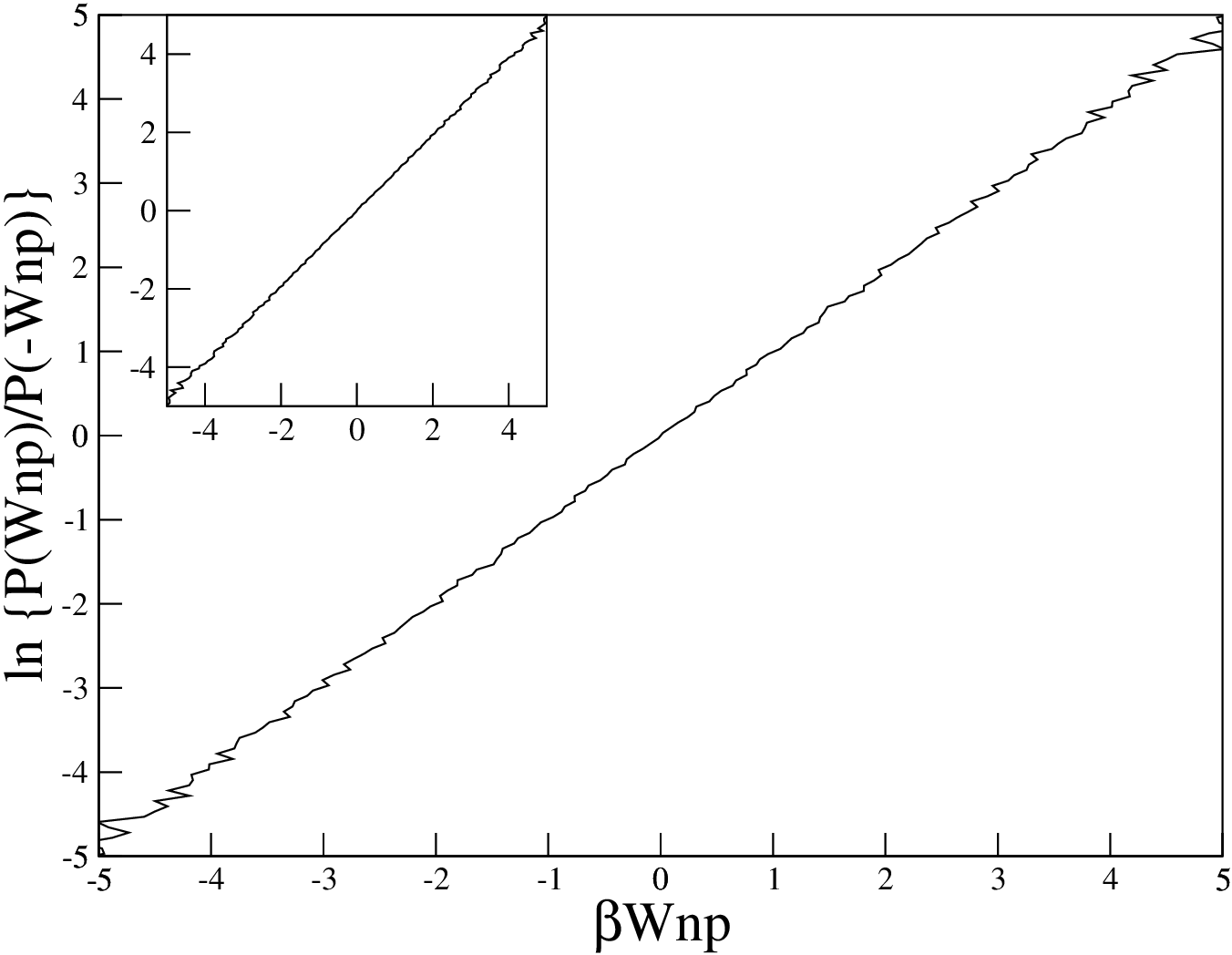}
\caption{\label{fig:epsart}Plot of ln$\{P(W_{np})/P(-W_{np})\}$ vs. 
$\beta W_{np}$ 
(over 10 periods). Inset shows the corresponding plot of $W_{np}$
(over 5 periods). The plots are for D=0.16,$ \omega=0.1$ and A=0.1}
\end{figure}

We see that $V_{r}>1$ in the parameter range we have chosen. 
This shows that fluctuations dominate 
the mean value. $V_r$ exhibits a minimum near the value of D at which 
SR occurs. It may be
noted that when $V_r$ becomes much larger than one the average value 
ceases to be a good 
physical variable (non self-averaging quantity) and hence to 
characterise the behavior of 
work one has to resort to the study of the entire distribution 
function. In the time periodic state the average work done is dissipated 
into the system as heat. Thus one can identify $\langle W_p\rangle$ as a 
hysteresis loss in the medium. However, it may be noted that fluctuations 
of work done cannot be identified with heat fluctuations \cite{Mamata1,Cohen}.

Now we will turn to the understanding of the nature of probability 
distribution $P(W_p)$ of $W_p$ for various values of noise strength 
D across SR peak. 
For this we have plotted $P(W_p)$ as a function of $W_p$in Figs. 2a and 2b 
for various values of D.  
For low values of D ($\sim$ 0.02) particle dynamics is mostly confined
to a small amplitude intrawell oscillatory motion.
Hence the distribution is closer to the Gaussian. For 
slightly higer values of D($\simeq$ 0.04 and 0.06) the particle makes 
occassional random excursions into the other well as a function of time. 
This is clearly reflected as a small asymmetry and a hump (interwell motion) in the 
plot of $P(W_p)$. It is interesting 
to note that there is substantial weight towards 
negative values of $W_p$. The negative values of $W_p$ corresponds to 
transient second law violating hysteresis loops. As we increase D further 
the interwell dynamics starts playing a prominent role.  
 It is interesting to note that position of the first 
and the second peak does not shift much, however, its width increases. 
The weight
 of the probability distribution towards higher negative values of $W_p$  
increases. On further increasing D (Fig.2b) 
finally a single peak structure appears. The shape of the 
single peak structure is closer to a Gaussian at high D($\sim$ 0.5). 
For such high values of D 
particle makes several random excursions between the two wells during a 
single time period of an external drive. Hence total work can be treated 
as an addition of independent increments of work. Then the central limit
theorem leads us to expect that the distribution of work will be 
approximately Gaussian. This is indeed the case. The parameter 
range for which we have explored the distributions are broad and the 
variance of 
fluctuations dominate over the average values (Fig.1, inset). This is one 
of our main observations.

Finally we discuss the applicability of SSFT for the work done over a 
single period. For this we have plotted in Fig.3, $P(W_p)$ and 
$P(-W_p)exp(\beta{W_p})$
for three different values of D. For small D ($\simeq$ 0.02), Fig.3a and 
for large D
($\simeq$ 0.5), Fig.3c $P(W_p)$ and $P(-W_p)exp(\beta W_p)$ are almost equal, 
thus satisfying SSFT for work distribution. For the intermediate value
of D ($\simeq$0.16), Fig.3b we see that SSFT does not hold\cite{Dhar}. This 
clearly indicates 
that prediction
of validity of SSFT for work done over a single period is restricted only to
a driven overdamped harmonic oscillator (linear problem)\cite{Mamata1}. 
This will not 
be the case in the presence of nonlinearity. We consider a parameter regime 
where $D = 0.6$, $\omega=0.01$, $A=0.1$ of Fig. 3, where SSFT over a 
single period does not hold. However, we calculate the work distribution 
$W_{np}$ for 5 and 10 periods. 

In Fig.4 we have plotted $ln(\frac{(P(W_{np})}{P(-W_{np})})$ as a function 
of $\beta W_{np}$ for periods 5 (inset) and 10. 
 From the plots it is obvious that the slope is 1 
and hence proving the validity of SSFT \cite{Douarche}. Moreover, the observed
probability distribution is Gaussian and its variance V is related to the mean 
value $\langle W_{np}\rangle$, i.e., 
$V=\frac{2}{\beta}{\langle W_{np} \rangle}$,
which is a requisite for SSFT if the distribution is Gaussian 
(eqn. 24 of\cite{Mamata1}). 

Thus we see that work 
distribution satisfies SSFT provided we consider work over a large number
of periods. The number of periods above which SSFT holds depends on the 
physical parameters. The finite time corrections are very 
complex\cite{Douarche} which we have not discussed here.

\section{Conclusion:}
We have studied a problem of work fluctuations in a driven double well 
system which exhibits SR.We have shown that across SR fluctuations of work 
calculated over a single period in the asymptotic time region dominates the 
mean value. The distributions exhibit significant weight towards 
negative values of work indicating transient second law violating 
trajectories. As opposed to the case of a linear problem of driven 
harmonic oscillators SSFT for work 
distribution is not satisfied for a work calculated over a single period. 
However, if the work is calculated over a large number of periods SSFT is 
indeed satisfied. We have also calculated the distributions of internal energy
and dissipated heat over a cycle so as to verify the extended form of SSFT
\cite{Cohen}. The work will be reported elsewhere.

\section{Acknowledgement:}
SS and RR gratefully acknowledges Institute of Physics, Bhubaneswar for 
hospitality. RR also wishes to thank DST for financial assistance 
(SR/FTP/PS-33/2004).

\end{document}